\documentclass[twocolumn,showpacs,preprintnumbers]{revtex4}
\usepackage{amsmath}
\usepackage[dvips]{graphicx}
\usepackage{bm}

\begin{document}

\title{Magnetic properties of bi-phase micro- and nanotubes}
\author{J. Escrig$^1$}
\author{D. Altbir$^1$}
\author{K. Nielsch$^2$}
\affiliation{$^{1}$Departamento de F\'{\i}sica, Universidad de Santiago de Chile, USACH,
Av. Ecuador 3493, Santiago, Chile\\
$^{2}$Max-Planck-Institut of Microstructure Physics, Weinberg 2, 06120
Halle, Germany}
\date{\today }

\begin{abstract}
The magnetic configurations of bi-phase micro- and nanotubes consisting of a
ferromagnetic internal tube, an intermediate non-magnetic spacer, and an
external magnetic shell are investigated as a function of their geometry.
Based on a continuum approach we obtained analytical expressions for the
energy which lead us to obtain phase diagrams giving the relative stability
of characteristic internal magnetic configurations of the bi-phase tubes.
\end{abstract}

\maketitle

\section{Introduction}

The fabrication of ultra-highly magnetic recording media has become one of
the great challenges in the field of magnetism. It is well known that in
conventional recording media, composed of weakly coupled magnetic grains,
the areal density is limited by the so-called superparamagnetic effect, \cite%
{Ross} which arises as the grain size is reduced below $10$ nm. In such a
case, thermal fluctuations in the magnetic moment orientation may lead to
loss of information. In order to increase the recording densities, new
magnetic elements need to be proposed.

Patterned magnetic media consisting of regular arrays of magnetic layered
nanodots \cite{Albrecht} or nanorings \cite{Escrig} have been considered as
providing the basis for extending magnetic storage densities beyond the
superparamagnetic limit. In such systems, a single dot or ring with $n$
magnetic layers, whose volume is much larger than those of the grains in
conventional recording media, might store up to 2$^{n}$ bits, beating
thermal fluctuations and increasing the recording density by a factor of 2$%
^{n-1}$.

Recently, magnetic nanotubes can be (chemically) functionalized on the in-
and outside, \cite{Nielsch, Nielsch2, Wang} motivating a new field of
research. Nanotubes exhibit a core-free magnetic configuration leading to
uniform switching fields, guaranteeing reproducibility \cite{Wang, Escrig2}
and have been proposed to be used as bi-directional sensors by Lee \textit{%
et al}. \cite{Lee} Also due to their low density they can float in
solutions, making them suitable for biological applications (see [4] and
references therein). Numerical simulations \cite{Wang, Landeros} and
analytical calculations \cite{Escrig2, Landeros} on such tubes have
identified two main ideal magnetic configurations: a ferromagnetic state
with all the magnetic moments pointing parallel to the axis of the tube ($c=1
$) and an in-plane magnetic ordering, namely the flux-closure vortex state ($%
c=2$). However, in tubes with low aspect ratio $(L/R\leq 10)$, another state
with two opposite directed vortices with a domain wall between them can
appear. \cite{Wang, Lee} Similar nano-objects, bi-phase microwires
consisting of two metallic layers, a cylindrical nucleus and an external
magnetic microtube, separated by an intermediate nonmagnetic microlayer,
have been introduced by Pirota \textit{et al}. \cite{Pirota} .

The synthesis of bi-phase microwires and magnetic nanotubes opens the
possibility of fabricating bi-phase micro- and nanotubes, where new magnetic
phases can appear. Because of the geometry, the internal and external tubes
can be close enough to interact via a strong dipolar coupling. This
interaction can produce new magnetic states of the cylindrical particle as a
whole, \cite{Xiao} which can be used for particular applications. Besides,
we can expect the appearance of new magnetic properties, like the dipolar
magnetic bias responsible for the giant magnetoimpedance behavior of
amorphous microwires. \cite{Sinnecker} The possibility of achieving such a,
spin-valve-like, hysteresis loop is very attractive because of potential
applications to sense magnetic fields in magnetic recording systems.

\begin{figure}[h]
\begin{center}
\includegraphics[width=8cm]{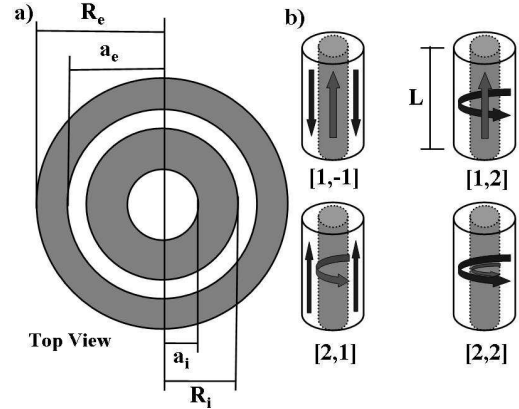}
\caption{ (a) Geometrical parameters and (b) characteristic magnetic
configurations of a bi-phase micro- and nanotube.}
\end{center}
\end{figure}

The purpose of this paper is to investigate the magnetic ordering of
bi-phase tubes. Our cylindrical particle, of length $L$, consists of two
magnetic tubes separated by a nonmagnetic spacer. The internal (external)
tube is characterized by its outer $R_{i}\left( R_{e}\right) $ and inner $%
a_{i}\left( a_{e}\right) $ radii, as illustrated in figure 1(a). Magnetic
configurations of the bi-phase tubes can be identified by two indices $%
[c_{i},c_{e}]$, where $c_{i},c_{e}=$ 1 or 2 denote the magnetic
configurations of the internal and external magnetic tubes, respectively
(see figure 1(b)).

\section{Total energy calculations}

We adopt a simplified description of the system, in which the discrete
distribution of magnetic moments is replaced by a continuous one
characterized by a slowly varying magnetization density $\vec{M}(\vec{r})$.
The total energy $E_{tot}^{[c_{i},c_{e}]}$ is generally given by the sum of
three terms, the magnetostatic $E_{dip}^{[c_{i},c_{e}]}$, the exchange $%
E_{ex}^{[c_{i},c_{e}]}$, and the anisotropy contributions. Here we are
interested in soft or polycrystalline magnetic materials, in which case the
anisotropy contribution is usually disregarded. \cite{Nielsch, Wang}

The total magnetization can be written as $\vec{M}(\vec{r})=\vec{M}%
_{i}\left( \vec{r}\right) +\vec{M}_{e}\left( \vec{r}\right) $, where $\vec{M}%
_{i}\left( \vec{r}\right) $ and $\vec{M}_{e}\left( \vec{r}\right) $ denote
the magnetization of the internal and external tubes, respectively. In this
case, the magnetostatic potential $U\left( \vec{r}\right) $ splits up into
two components, $U_{i}\left( \vec{r}\right) $ and $U_{e}\left( \vec{r}%
\right) $, associated with the magnetization of each individual tube. Then,
the total dipolar energy can be written as $%
E_{dip}^{[c_{i},c_{e}]}=E_{d}^{[c_{i}]}+E_{d}^{[c_{e}]}+2E_{d}^{[c_{i},c_{e}]},
$ where $E_{d}^{[c_{p},c_{q}]}=\left( \mu _{0}/2\right) \int \vec{M}%
_{p}\left( \vec{r}\right) \nabla U_{q}\left( \vec{r}\right) \,dv$, with $%
p,q=i,e$. The diagonal terms, $E_{d}^{[c_{p},c_{p}]}\equiv E_{d}^{[c_{p}]}$,
correspond to the dipolar contributions to the self-energy of the individual
magnetic tubes, whereas the off-diagonal term, $E_{d}^{[c_{p},c_{q}]}$,
corresponds to the dipolar interaction between the two tubes.

The exchange energy $E_{ex}^{[c_{p},c_{q}]}$ also has three contributions,
two coming from the direct exchange interaction within the magnetic tubes,
and the other one from its indirect interaction mediated by the non-magnetic
spacer. We focus on systems in which the thickness of the intermediate tube,
$d_{s}=a_{e}-R_{i}$, is large enough to make negligible the indirect
exchange interaction between the two magnetic tubes across the non-magnetic
one. A good estimate of the range of the indirect exchange interaction can
be obtained from results for multilayers.\cite{Bloemen} In general the
interlayer exchange coupling vanishes for spacer thicknesses greater than a
few nanometers. Here we focus our attention on those cases in which $d_{s}$
is bigger than $5$ nm, thus, to a good approximation, $%
E_{ex}^{[c_{p},c_{q}]} $ can be written as $%
E_{ex}^{[c_{i}]}+E_{ex}^{[c_{e}]} $, with $E_{ex}^{[c_{p}]}=A\int \left[
\left( \nabla m_{px}\right) ^{2}+\left( \nabla m_{py}\right) ^{2}+\left(
\nabla m_{pz}\right) ^{2}\right] dv,$ \cite{Aharoni} $\vec{m}%
_{p}=(m_{px},m_{py},m_{pz})=\vec{M}_{p}/M_{0}$ is the magnetization of tube $%
p$ normalized to the saturation magnetization $M_{0}$ and $A$ is the
stiffness constant of the magnetic material.

The total energy of the multilayered structure can be written in the form $%
E_{tot}^{[c_{i},c_{e}]}=E_{self}^{[c_{i}]}+E_{self}^{[c_{e}]}+E_{int}^{[c_{i},c_{e}]}
$, where $E_{self}^{[c_{p}]}=$ $E_{d}^{[c_{p}]}+E_{ex}^{[c_{p}]}$ is the
self-energy of each magnetic tube, and $E_{int}^{[c_{i},c_{e}]}=$ $%
2E_{d}^{[c_{i},c_{e}]}$ is the (dipolar) interaction energy between them. We
now proceed to calculate the energy terms in the expression for $%
E_{tot}^{[c_{1},c_{2}]}$. Results are given in units of $\mu
_{0}M_{0}^{2}l_{ex}^{3}$, i.e, $\tilde{E}=E/\mu _{0}M_{0}^{2}l_{ex}^{3}$,
where $l_{ex}=\sqrt{2A/\mu _{0}M_{0}^{2}}$.

In order to perform the integrals in the above expressions, it is necessary
to specify the functional form of the magnetization density for each
configuration. For $c=1$, $\vec{M}_{p}\left( \vec{r}\right) $ can be
approximated by\ $M_{0}\hat{z}$, where $\hat{z}$ is the unit vector parallel
to the axis of the tube. In this case, we find that the reduced
self-energies take the form%
\begin{equation*}
\tilde{E}_{self}^{[1]}(p)=\frac{\pi R_{p}^{2}}{l_{ex}^{3}}%
\int\limits_{0}^{\infty }\frac{dq}{q^{2}}\left( 1-e^{-qL}\right) ~\left(
J_{1}(qR_{p})-\beta _{p}J_{1}(qa_{p})\right) ^{2},
\end{equation*}%
with $J_{1}(z)$ a Bessel function of first kind. In this case, the exchange
contribution to the self-energy vanishes. For $c=2$, $\vec{M}_{p}\left( \vec{%
r}\right) $ can be approximate by $M_{0}\hat{\phi}$, where $\hat{\phi}$ is
the azimuthal unit vector. In such case the dipolar contribution to the
self-energy turns out to be equal to zero, and $\tilde{E}_{self}^{[2]}(p)=-%
\left( \pi L\ln \beta _{p}\right) /l_{ex}$. We have defined $\beta
_{p}=a_{p}/R_{p}$.

Regarding the interaction between the magnetic tubes, the only non-zero
terms correspond to the case in which both tubes are in the $c=1$
configuration. Due to the condition of perfect flux closure in the vortex
configuration, a magnetic tube in such a configuration does not interact
with another, independently of the magnetic configuration of the latter.
Thus, we end up with just
\begin{multline*}
\tilde{E}_{int}^{\left[ 1,-1\right] }=-\frac{2\pi R_{i}R_{e}}{l_{ex}^{3}}%
\int\limits_{0}^{\infty }\frac{dq}{q^{2}}\left( 1-e^{-qL}\right)  \\
\left( J_{1}(qR_{i})-\beta _{i}J_{1}(qa_{i})\right) \left(
J_{1}(qR_{e})-\beta _{e}J_{1}(qa_{e})\right) \ .
\end{multline*}%
The minus sign in the superscript of $\tilde{E}_{int}^{[1,-1]}$ indicates
that the relative orientation of the magnetization of the internal and
external tubes is antiparallel. We remark that, due to the dipolar
interaction between the tubes, the total energy of the $\left[ 1,1\right] $
configuration is always larger than the total energy of $\left[ 1,-1\right] $%
.

\section{Results and discussion}

We are now in position to investigate the relative stability of the various $%
[c_{i},c_{e}]$ configurations. We consider nickel ($l_{ex}=8.225$ nm) and
cobalt ($l_{ex}=2.849$ nm) magnetic bi-phase tubes. However, to properly
describe this geometry we need to deal with five different geometrical
parameters. In order to simplify the illustration of our results, we present
phase diagrams in terms of $d_{i}/d_{e}$, where $d_{i}=R_{i}-a_{i}$ is the
thickness of the internal magnetic tube and $d_{e}=R_{e}-a_{e}$ represents
the thickness of the external magnetic shell, as illustrated in figure 2.
\begin{figure}[h]
\includegraphics[width=8cm]{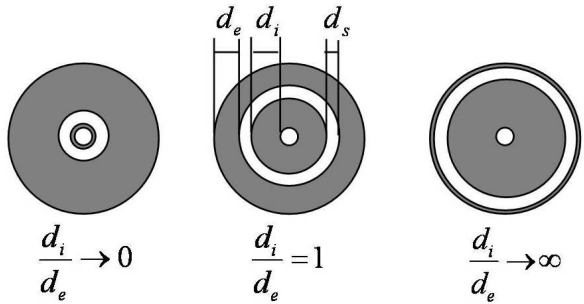}
\caption{ Geometrical parameters and limits of a bi-phase micro-
and nanotube. }
\end{figure}

We start by investigating systems with fixed $a_{i}=10$ nm and $d_{s}=50$
nm. Figure 3 illustrates the phase diagrams, giving the regions within which
one of the $[c_{i},c_{e}]$-configurations is of lowest energy for different
nanotube lengths. The diagrams show the existence of only three phases,
corresponding to the configurations $[1,-1]$, $[1,2]$, and $[2,2]$. The
absence of the $\left[ 2,1\right] $ phase can be understood from the work by
Escrig \textit{et al}. \cite{Escrig} By inspecting the phase diagram for
magnetic tubes presented in that work we can see that, for fixed length, it
is easier to find a vortex state for bigger external radius. Then it will be
much difficult to find a bi-phase tube with the inner tube in the vortex
phase and the external tube, which has a bigger radius than the internal
one, in the ferromagnetic phase. The $\left[ 2,1\right] $ phase is even less
probable in this interacting regime, because the dipolar interaction between
the tubes will favor the antiferromagnetic-like coupling. By inspecting the
energy equation we observe that the $[2,1]$ phase appears only in the limit $%
d_{i}/d_{e}\rightarrow \infty $. Besides, we observe that a shorter bi-phase
tube favors the appearance of vortex configurations. It is worth mentioning
that in our model the $[2,-2]$ configuration, corresponding to a vortex and
an anti-vortex, and the $[2,2]$ one have the same energy, since in both
cases the magnetic nanotubes do not interact. In what follows we shall refer
to both as $[2,2]$-configurations. By comparing our results for Ni and Co we
observe differences in the behavior of the triple point as a function of $L$%
. By looking to phase diagrams for isolated tubes, \cite{Escrig} we observe
that a vortex phase appears at smaller external radii of a Co tube than in a
Ni one because $l_{ex}(Co)\approx $ $3l_{ex}(Ni)$. This effect is also
present in bi-phase tubes, where the triple point occurs for smaller $R_{e}$
in Co. Because $l_{ex}$ is bigger in Ni than in Co, by varying the geometry
of the system, the triple point will experience stronger effects in Co than
in Ni.
\begin{figure}[h]
\includegraphics[width=8cm]{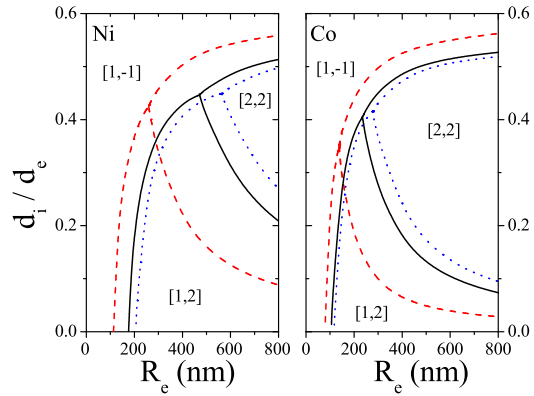}
\caption{(color online) Variable tube length. Phase diagrams for bi-phase
nanotubes with $a_{i}=10$ nm, $d_{s}=50$ nm and $L=1$ $\protect\mu $m
(dashed lines), $6$ $\protect\mu $m\ (solid lines), and $10$ \ $\protect\mu $%
m (dotted lines).}
\end{figure}
\begin{figure}[h]
\includegraphics[width=8cm]{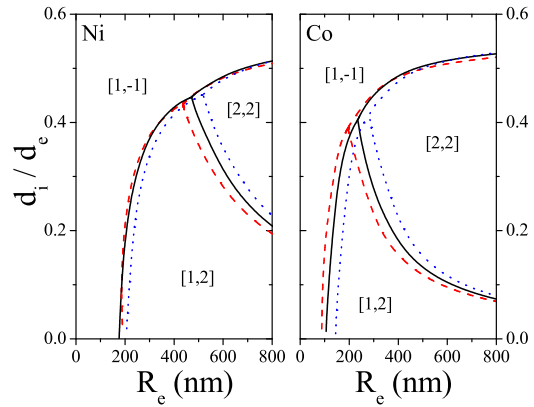}
\caption{(color online) Variable thickness of the spacer layer. Phase
diagrams for bi-phase nanotubes with $a_{i}=10$ nm, $L=6$ $\protect\mu $m
and $d_{s}=5$ nm (dashed lines), $50$ nm\ (solid lines), and $100$ \ nm
(dotted lines).}
\end{figure}

The dependence of the whole diagram on $d_{s}$ is illustrated in figure 4.
In a first inspection of the figures it seems that the main role in this
case is played by the material specified in our calculations by $l_{ex}$.
However it is important to recall that the ordinate is given by a composed
parameter, which makes it complex to determine from the figure the magnetic
behavior of the system. Because the interaction between the tubes
contributes to stabilize the $[1,-1]$ phase, the decay of the dipolar
interaction between the tubes as a function of $d_{s}$ increases the energy
of the $[1-1] $ phase, leading to the appearance of the other phases. We
illustrate this point with an example. Let us consider a bi-phase Ni tube
defined by $a_{i}=10$ nm, $d_{i}=30$ nm, $d_{e}=150$ nm and $L=6$ $\mu $m.
In this case $d_{i}/d_{e}=0.2$. If the cylindrical particle is defined by $%
d_{s}=5$ nm, $R_{e}=195$ nm, and from figure 4 we observe that the
nano-object will exhibit a $[1,-1]$ phase. But if for the same geometry we
chose $d_{s}=100$, then $R_{e}=290 $ and the cylindrical particle will be in
the $[1,2]$ phase.

We now turn our attention to the case in which $a_{i}$ is the variable (see
figure 5). Here, a strong shift of the transition lines is evidenced as
varying the internal radius of the nano-object. In particular, for
geometries defined by parameters inside the dark region, the system can
exhibit any of three ideal phases, depending on the value of $a_{i}$. Then a
clear determination of $a_{i}$ is fundamental when a particular magnetic
phase is searched for. In these figures, a clear predominance of the $[1,-1]$
phase appears while increasing $a_{i}$. By increasing $a_{i}$ with fixed $%
R_{e}$, $d_{s}$ and $d_{i}/d_{e}$, $R_{i}$ and $a_{e}$ necessarily increase.
In this way, both tubes have less cross-section area, and then the system
prefers the ferromagnetic phase $[1,-1]$.
\begin{figure}[h]
\includegraphics[width=8cm]{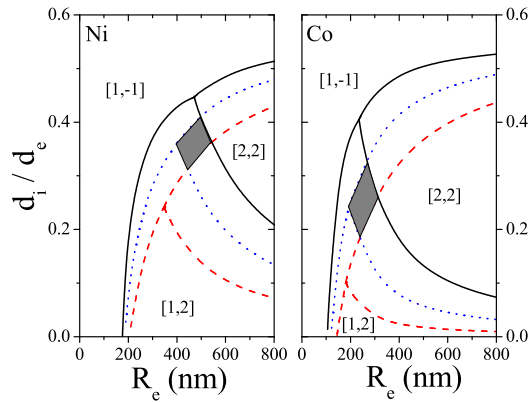}
\caption{(color online) Variable internal radius $a_{i}$. Phase diagrams for
bi-phase nanotubes with $d_{s}=50$ nm, $L=6$ $\protect\mu $m and $a_{i}=$ $%
10 $ nm\ (solid lines), $30$ \ nm (dotted lines), and $60$ nm (dashed
lines). Inside the dark region, the system can exhibit any of three ideal
phases, depending on the value of $a_{i}$.}
\end{figure}
\begin{figure}[h]
\includegraphics[width=8cm]{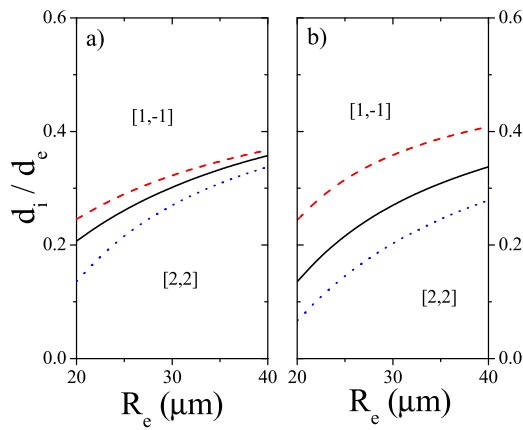}
\caption{(color online) Phase diagrams for bi-phase microtubes with a) $%
a_{i}=5$ $\protect\mu $m, $L=1$ cm and $d_{s}=$ $1$ $\protect\mu
$m\ (dashed lines), $5$ \ $\protect\mu $m (solid lines), and $10$
$\protect\mu $m (dotted lines) and b) $d_{s}=$ $5$ $\protect\mu
$m, $L=1$ cm and $a_{i}=$ $3$ $\protect\mu $m\ (dashed lines), $5$
\ $\protect\mu $m (solid lines), and $7$ $\protect\mu $m (dotted
lines).} 
\end{figure}

We also look to the phase diagrams of bi-phase microtubes. In this case, for
the range of diameters and lengths considered, the transition lines are
almost independent of the magnetic material. Our results are summarized in
figure 6, describing the phases for any magnetic material. Figure 6(a)
illustrates the dependence of the phase diagram on $d_{s}$, and figure 6(b)
shows the behavior of the whole diagram with $a_{i}$. The diagrams present
only two regions, corresponding to the phases $[1,-1]$ and $[2,2]$. The $%
[1,2]$ phase is suppressed for diameters in the range of the micrometers, as
expected from previous figures. Pirota \textit{et al. \cite{Pirota} }and V%
\'{a}zquez \textit{et al.} \cite{Vazquez} have investigated the magnetic
behavior of different multilayer microwires. Hysteresis loops clearly
exhibit two Barkhausen jumps, each related to the reversal of one of the
tubes. For small values of $d_{i}/d_{e}$, they observe that these Barkhausen
jumps are smooth, showing the existence of non-homogeneous states during the
reversal of each tube. This qualitative observation is in agreement with our
phase diagrams, which show that the $[2,2]$ configurations, which exhibit
non-homogeneous reversion, appear preferably for small $d_{i}/d_{e}$. By
comparing our results for nano- and microtubes we observe that the $[2,2]$
phase appears preferably in cylindrical particles with large diameters.

In our calculations we considered soft or polycrystalline materials and then
anisotropy was disregarded. However, in crystalline materials anisotropy
plays a fundamental role, as shown by Escrig \textit{et al.} \cite{Escrig3}
In fact, the existence of a strong uniaxial anisotropy, as in Co, favors the
$[1,-1]$ phase, decreasing the other two phases, specially the $[2,2]$ one.
Nickel has a weak cubic anisotropy with $K<0$, which will enhance the $[2,2]$
phase.

\section{Conclusions}

In conclusion, we have studied the relative stability of ideal
configurations of magnetic bi-phase micro- and nanotubes composed by an
internal magnetic tube, an intermediate non-magnetic spacer and an external
magnetic shell. In such systems we investigated the size range of the
geometric parameters for which different configurations are of lowest
energy. Results are summarized in phase diagrams which clearly indicate that
the magnetic behavior of such structures can be tailored to meet specific
requirements provided a judicious choice of such parameters is made. This
includes not just the control over the number of magnetic configurations the
system might exhibits, but also the possibility of suppressing specific
phases. The lines separating the magnetic phases and, in particular, the
triple point, are very sensitive to the geometry of the bi-phase tube.
Results in nanotubes strongly depend on the material, defined by its
exchange length. However, in microtubes the [1,2] phase is suppressed and
the phase diagrams are almost independent of the material. For magnetic
bi-phase tubes with diameter of less than 100 nm, the $[1,-1]$ configuration
is always present. Besides, we can conclude that for multilayer tubes where
total $M_{i}\geq M_{e}$ we always find the antiferromagnetic alignment. The
phase diagrams presented can provide guidelines for the production of
nanostructures with technological purpose such as the fabrication of sensor
devices.

\begin{acknowledgments}
This work has been partially supported in Chile by FONDECYT 1050013,
Millennium Science Nucleus Condensed Matter Physics P02-054F and a CONICYT
PhD Program Fellowship. Financial support from the German Federal Ministry
for Education and Research (BMBF, Project No 03N8701) is gratefully
acknowledged. JE is grateful to the Max-Planck-Institut for Microstructure
Physics, Halle, for its hospitality. We thank M Daub and P Landeros for
useful discussions.
\end{acknowledgments}

\end{document}